\documentclass[twocolumn,footinbib,floatfix,aps,10pt,prl]{revtex4-1}
\usepackage{graphicx}
\usepackage{epsfig}
\usepackage{epstopdf}
\usepackage{amsmath}
\usepackage{amsfonts}
\usepackage{txfonts}
\usepackage{amssymb}
\usepackage{amstext}
\usepackage{rotating}
\usepackage{dcolumn}
\usepackage{graphics}
\usepackage{array,multirow}
\usepackage[english]{babel}
\usepackage{attachfile}
\usepackage{pgfplots}
\usepackage[export]{adjustbox}
\usepackage{braket}
\usepackage{hyperref}

\def\infinity{\rotatebox{90}{8}}
\def\be{\begin{equation}}
\def\ee{\end{equation}}
\def\bea{\begin{eqnarray}}
\def\eea{\end{eqnarray}}

\newcommand{\mbp}[1]{{\color{black} #1}}
\newcommand{\lp}{\left(}
 \newcommand{\rp}{\right)}

\begin{document}

\title{Phase-dependent exciton transport and \emph{energy harvesting} from thermal environments}

\author{S. Oviedo-Casado${}^{1}$, J. Prior${}^{2}$, A.W. Chin${}^{3}$, R. Rosenbach${}^{1}$, S.F. Huelga${}^{1}$, and M.B. Plenio${}^{1}$}
\affiliation{${}^{1}$ Institute of Theoretical Physics, Albert-Einstein-Allee 11, D - 89069 Ulm, Germany\\
${}^{2}$ Departamento de F{\'i}sica Aplicada, Universidad Polit{\'e}cnica de Cartagena, Cartagena 30202 Spain\\
${}^{1}$ Theory of Condensed Matter Group, Cavendish Laboratory, University of Cambridge, Cambridge, CB3 0HE, UK}

\begin{abstract}
Non-Markovian effects in the evolution of open quantum systems have recently \mbp{attracted} widespread
interest, particularly in the context of assessing the efficiency of energy and charge transfer in
nanoscale biomolecular networks and quantum technologies. With the aid of many-body simulation methods,
we uncover and analyse an ultrafast environmental process that causes energy relaxation in the reduced
system to depend explicitly on the phase relation of the initial state preparation. Remarkably, for
particular phases and system parameters, the net energy flow is uphill, transiently violating the
principle of detailed balance, and implying that energy is spontaneously taken up from the environment.
A theoretical analysis reveals that non-secular contributions, \mbox{significant only} within the environmental
correlation time, \mbp{underlie} this effect. This suggests that environmental energy harvesting will be observable
across a wide range of coupled quantum systems.
\end{abstract}

\maketitle
%


In recent years, new trends have \mbp{emerged} in the field of open quantum system
dynamics, particularly in the general area of ultrafast, non-Markovian dynamics
\cite{Rivas2014} \mbp{for which the
primary steps of natural photosynthesis provide a fascinating testing ground.}
\mbp{In this area} the observation of \mbp{surprisingly long-lasting coherence} in a broad
range of photosynthetic pigment-protein complexes (PPCs)
\cite{engel2007evidence,collini2010coherently,Engel2010,vanHulst2013,romero2014quantum,Fuller2014}
as well as other coherence effects \cite{gelinas2014ultrafast,Ghosh2015}
has generated tremendous interest in understanding whether quantum effects may underpin the
near-unit efficiency of the processes of exciton transport and charge separation \mbp{(see
\cite{HuelgaP13} for a recent review)}. Early studies of PPCs, such as the Fenna-Matthews-Olson
(FMO) complex \cite{PlenioHuelga2008,mohseni2008environment,olaya2008efficiency,caruso2009highly, rebentrost2009environment, adolphs2006proteins,ishizaki2009theoretical},
have already established that a dynamical \emph{interplay} of coherent and dissipative
dynamics optimizes targeted exciton transfer. \mbp{Subsequent work has developed further}
the theme that \emph{dissipative} quantum dynamics may promote the efficiency of tasks in
photosynthetic and other organic light-harvesting materials, with a particular focus
on the \emph{complex, structured environments} often found in supramolecular systems
\cite{prior2010efficient,chin2012coherence,rey2013exploiting,nalbach2010quantum, kolli2012fundamental,creatore2013efficient,dorfman2013photosynthetic,Olaya2014,nathan2014}.
Motivated by the pressing need to simulate \mbp{accurately system-environment dynamics}
beyond the Markovian regime, a range of advanced open-system techniques have been developed
\cite{ritschel2011efficient,prior2010efficient,chin2010exact,kreisbeck2012long,nalbach2010quantum}.
Of these, the TEDOPA (Time Evolving Density with Orthogonal Polynomial Algorithm) method
has emerged as one of the most powerful for the characterization of transient dynamics
\cite{prior2010efficient,chin2010exact} \mbp{with rigorous error bounds \cite{WoodsCP15}}.
Due to its unique ability to track the many-body entangled state of \emph{both} the system
and its macroscopic environment, TEDOPA has shed new light into the mechanics of non-equilibrium
open dynamics in a range of molecular PPCs, solid state and abstract dissipative systems
\cite{prior2010efficient,chin2011generalized,prior2013quantum}.

Here we study a biologically motivated model system, namely, a molecular dimer \mbp{dominating}
the lowest energy exciton in the FMO aggregate, and use finite-temperature TEDOPA to explore
the dynamics within the correlation (memory) time $\tau_c$ of the environment, a regime in
which \mbp{simple master equation approaches tend to fail} \cite{BreuerPetruccione, RivasHuelga}.
Most strikingly, we find that the rate and even \emph{direction} of energy transfer becomes
dependent on the phase of \mbp{the initial} superposition of exciton states, allowing certain
preparations to extract energy \mbp{during} the non-equilibrium evolution of the environment.
With the aid of a time local master equation (ME), we pinpoint the physical origin of this
effect \mbp{to} non-secular contributions \mbp{of} the non-Markovian dynamics of the excitonic
system due to the quantum interference of dephasing (transversal) and relaxation (longitudinal)
fluctuations. In many situations, these effects will average out \mbp{over very short time scales}.
However, the dipolar coupling between the highly absorbing pigments in a densely packed PCC
\mbp{realises} an excitonic landscape \mbp{whose} energy splittings are typically \mbp{such that}
the secular time of the excitonic system can become comparable to the environmental correlation
time. \mbp{As a result} non-secular terms are no longer negligible. Our analysis not only allows
us to rationalise the trends seen in the TEDOPA data across the parameter space but in addition
demonstrates how the energy harvesting process can occur in a much broader range of open quantum
systems. Non-secular contributions have been recently shown to increase the degree of non-Markovianity
\cite{Mottonen2013} and have been studied in the context of single driven systems, where an enhanced
secular time emerges as a result of dressing by the external field \cite{Sabrina2010}. Finally, we
suggest how this effect could be observed in biomolecular or artificial devices.

\begin{figure}[!h]
\includegraphics[width=8.5cm]{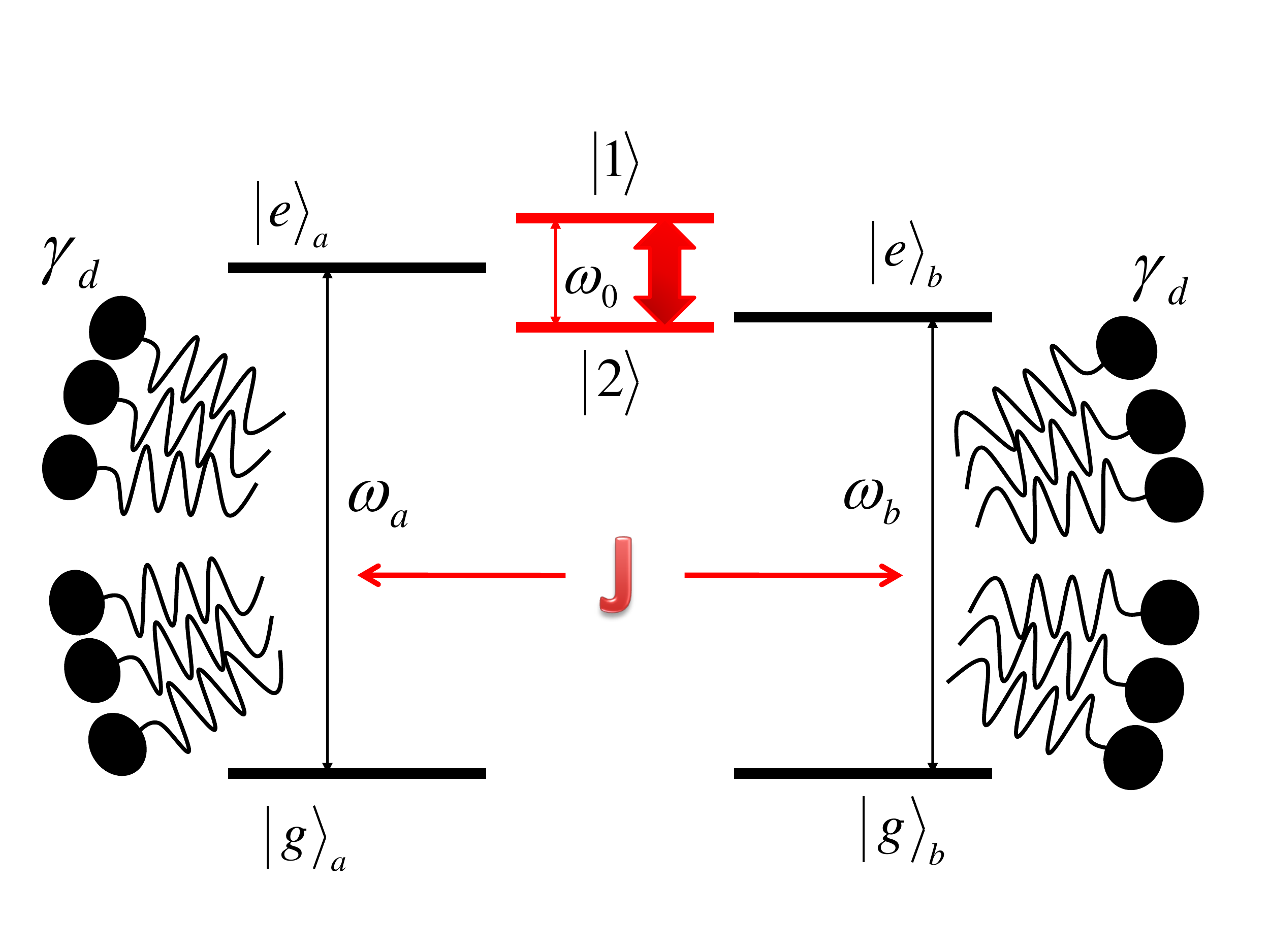}
\caption{Schematic representation of a dimeric model system subject to local dephasing $\gamma_d$.
Coherent inter-site electronic coupling $J$ yields the formation of excitonic eigenstates (states
$|1\rangle$ and $|2\rangle$). For $J = 52 \, cm^{-1}$ and optical frequencies $\omega_a$ and $\omega_b$,
the excitonic splitting $\omega_0$ falls in the far infrared. Exciton dynamics is dictated by transverse
(dephasing) and longitudinal (relaxation, marked with a red double arrow), environmental processes, whose
relative strength is governed by the exciton delocalization over sites. In the ultrafast regime, these
two types of fluctuation may interact, which leads to long lasting non-secular effects.  }
\label{Figure1}
\end{figure}

$\textit{The model}$ - Let us begin by describing a model dimer system as composed by two pseudo-spin-1/2
particles (pigment sites $a$ and $b$ with transition frequency $\omega_{a,b}$ (see Fig. \ref{Figure1}))
coupled via an exchange interaction of strength $J$. The system Hamiltonian reads $\mathcal{H}_S =
\frac{\omega_a}{2}\sigma_z^a + \frac{\omega_b}{2}\sigma_z^b + J\left(\sigma_+^a\sigma_-^b +
\sigma_-^a\sigma_+^b\right)$, where $\sigma_j$ $(j=x,y,z)$ are the standard Pauli matrices.
Each site is subject to the action of a dephasing (transversal) environment that produces phase
randomization but leaves site populations unaffected. We model this local interaction by coupling
the system to a continuum of harmonic oscillators $\mathcal{H}_B = \sum_{i} \int_0^{\infinity} dk h_i(k)
b_i^{\dagger}(k)  b_i(k)$, where the independent bosonic operators obey $\left[ b_k^i, b_l^{j\dagger}\right]
= \delta_{k,l}\delta_{i,j}$, via the Hamiltonian $\mathcal{H}_I = \sum_i \lp \sigma_z^i + I \rp/2
\int_0^{\infinity} dk g_i(k) \lp  b_i(k) +  b_i^{\dagger}(k) \rp $. Without loss of generality we will
assume the same dispersion relation $h(k)$ and coupling strength $g(k)$ for each site.

The coherent electronic coupling $J$ leads to the formation of delocalised eigenstates (excitons).
We restrict the excitonic manifold to the diagonalization of the \mbp{single}-excitation sector of
$\mathcal{H}_S$, spanned by $\ket{e_ag_b} = \ket{e}_a\otimes\ket{g}_b,\ket{g_ae_b} = \ket{g}_a\otimes\ket{e}_b$,
on the basis of biomolecules usually being subject to weak external illumination and/or by doubly excited
states typically being strongly suppressed \cite{Renger2001}. The exciton energies are then $E_{1,2} =
\pm\frac{\omega_{0}}{2}$ with $\omega_{0} = \sqrt{4J^2 + \Delta^2}$, corresponding to excitonic eigenvectors
\be
    \left(\begin{array}{c}
        \ket{1} \\
        \ket{2}
    \end{array} \right)
    =
    \left(\begin{array}{cc}
        cos\frac{\theta}{2} & sin\frac{\theta}{2} \\
        -sin\frac{\theta}{2} & cos\frac{\theta}{2}
    \end{array} \right)
    \left(
    \begin{array}{c}
        \ket{e_ag_b} \\
        \ket{g_ae_b}
    \end{array}\right),
\label{basis}
\ee
where $\Delta = \omega_a - \omega_b \, \lp\omega_a > \omega_b\rp$, and $\theta$ \mbp{denotes the}
\emph{mixing angle}, defined through the relation $tan \,\theta = 2J/\Delta$ with $0\leq\theta\leq\pi/2$.

Using TEDOPA we study the (exact) time evolution of an excitonic superposition state of the form
$\ket{\psi} = \frac{1}{\sqrt{2}}\lp e^{i\xi}\ket{1} +  \ket{2}\rp$, with a controllable phase $\xi$.
For concreteness, we choose a parameter regime as defined by the model system provided by sites $3$
and $4$ in the seven-site FMO Hamiltonian of $\it{C.\, Tepidum}$ as taken from \cite{adolphs2006proteins,chin2013role},
yielding a mixing angle $\theta \approx \pi/4$. The local environment is characterised by the smooth
part of the experimentally fitted super-ohmic spectral function of Adolphs and Renger (AR)
\cite{adolphs2006proteins} for the FMO complex. The reorganisation energy is 35 $cm^{-1}$ and
the background modes are assumed to be initially in thermal equilibrium at temperature $T$. Typical
values for the spectral width yield $\tau_c \sim 600$ fs while the system's secular time is $\omega_0^{-1}\sim 200$ fs.
The time scale over which non-secular effects manifest can therefore be greatly enlarged as compared
to isolated pigments.

\begin{figure}[!b]
    \includegraphics[width=7cm]{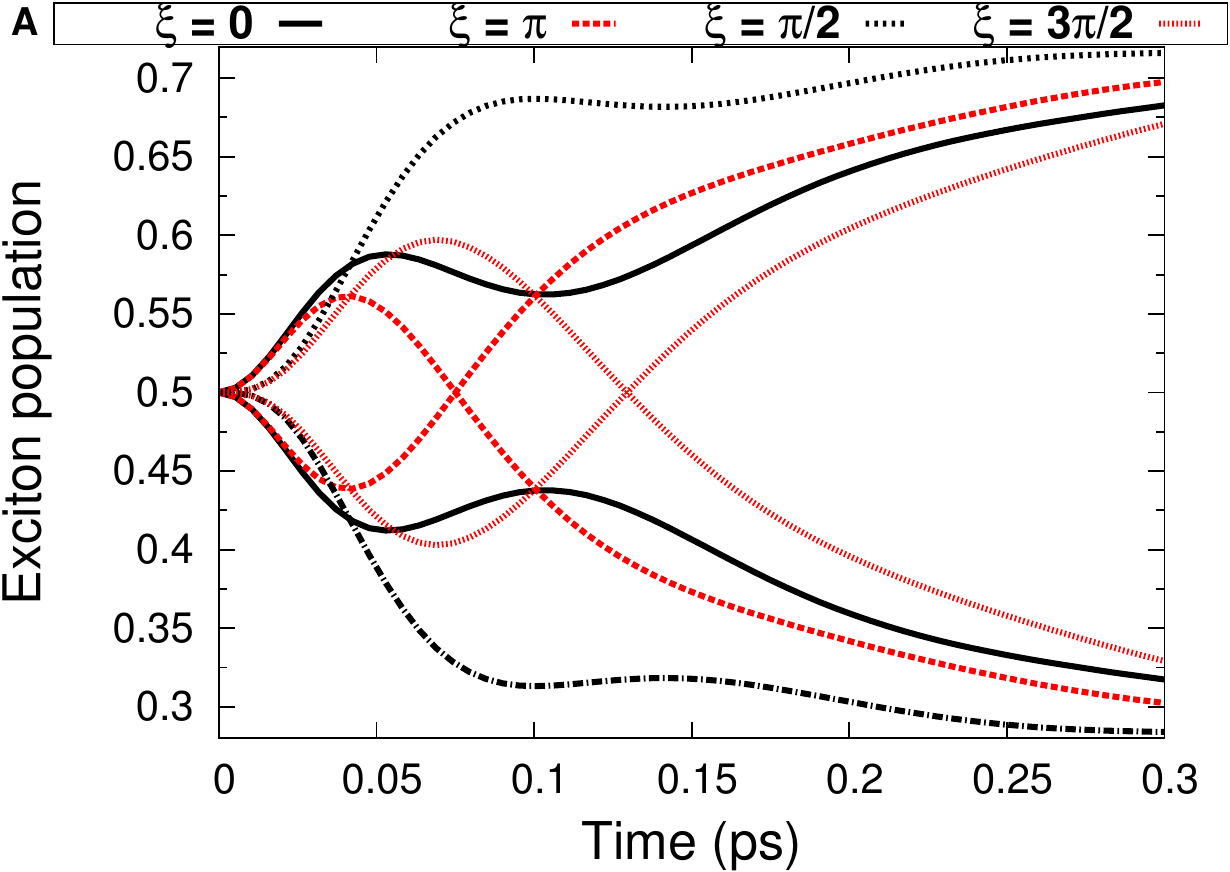}
    \includegraphics[width=7cm]{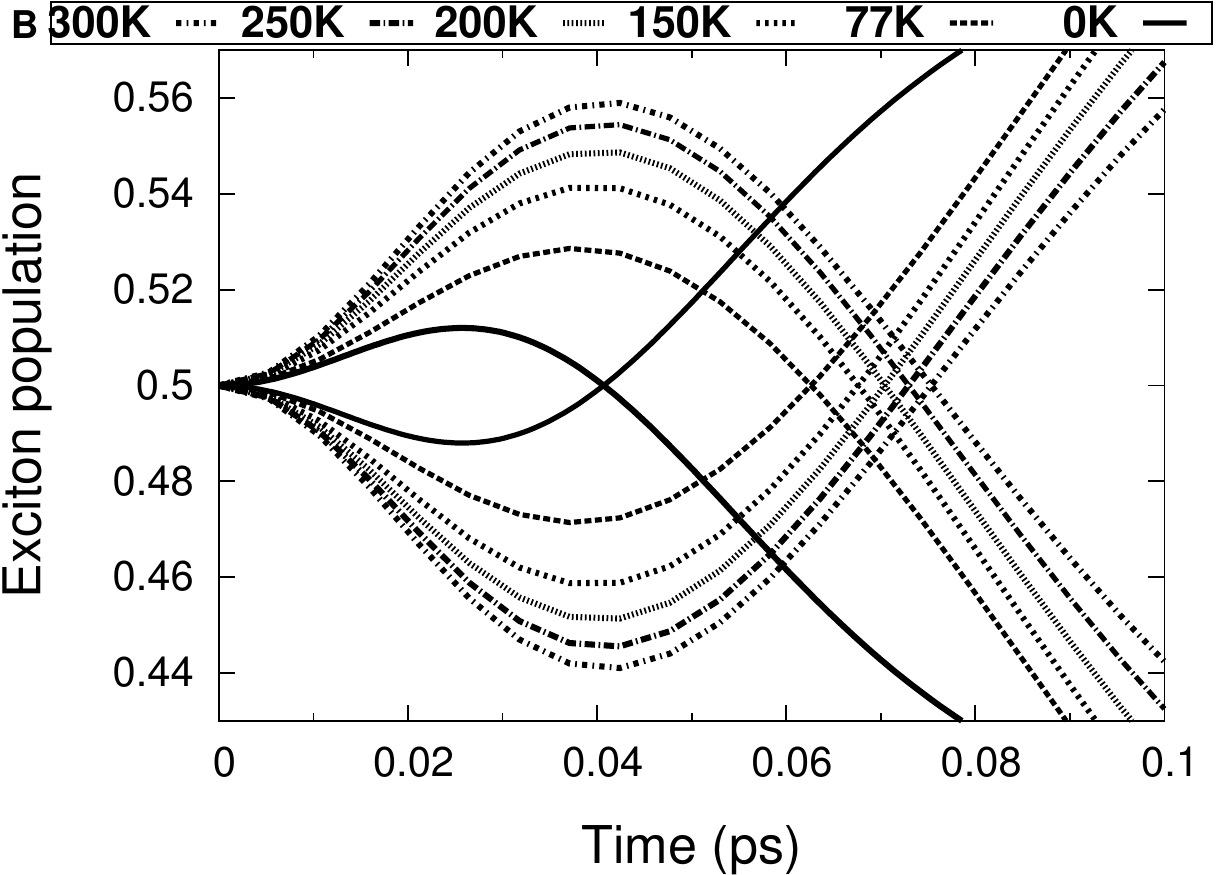}
    \caption{(Exact) Time evolution of excitonic populations as calculated by TEDOPA for a dimer
    system with $\omega_a -\omega_b = 135 \, cm^{-1}$ and $J = 52 \, cm^{-1}$ for representative
    $\xi = 0, \pi/2, \pi, 3\pi/2$  phases at $T = 277$ K in (A) and for temperatures $0-300$K
    following the initial state preparation with $\xi = \pi$ in (B). In both cases the system is
    subject to the super-ohmic dephasing background characterized by the AR spectral density \ref{SI}.}
    \label{Figure2}
\end{figure}

Figure \ref{Figure2}(a) shows the time evolution of the excitonic populations at T = 277 K for
four different initial superposition states with different phase $\xi$. We observe that while
for an initial phase $\xi = 0$ or $\pi/2$ the system relaxes monotonically to the equilibrium
state, with energy being continuously transferred from the system into the environment, the
behaviour for an initial phase of $\xi = \pi$ or $3\pi/2$ yields differs radically in the early
time evolution, with the population of the high energy exciton $\ket{1}$ becoming \emph{larger}
(population inversion) than the population of the low energy exciton for $\approx 100$ fs. These
results suggest that for a phase chosen \mbox{in} one half of the complex plane, energy flows
from the environment into the system at the early times, although the subsequent evolution is a
relaxation towards a unique equilibrium state where only the lowest energy exciton is populated.

In Figure \ref{Figure2}(b) we set the initial phase to be $\xi = \pi$, \mbp{allowing for energy
extraction from the environment,}
and study the excitonic dynamics for different values of the temperature. Interestingly, increasing
the temperature enhances the effect, inducing a stronger population inversion and extending the
duration of the anomalous dynamics before monotonic
relaxation sets in.

$\textit{Theory}$ - A formal derivation \mbp{of} a ME that ensures evolution under a dynamical
semigroup (commonly referred as a Lindblad ME), applied to a non-degenerated system, decouples
the evolution of coherences and populations in the \mbp{eigenbasis of the Hamiltonian} \cite{AlickiLendi}.
Therefore it will not capture any initial phase effect on the exciton population dynamics.
Moreover, in a Lindblad ME (or a secular Redfield ME) for a two level system, the excitonic
populations obey a Pauli ME \cite{PauliME}, implying monotonicity in the evolution at all
times, thence forbidding a population inversion as observed in Figs. \ref{Figure2}(a) and
(b). Both statements imply that to gain physical insight about the population inversion
witnessed by TEDOPA we need to consider a more general microscopic approach. Here, we work
in the Born approximation through the Redfield equation \cite{Redfield1957,Blum1981} to analyse
bath-mediated couplings between populations and coherences. In the exciton basis, we obtain
a time-local ME leading to a coupled set of differential equations for the evolution of
excitonic populations and coherences of the form
\be
    \label{reducedequations}
    \begin{array}{c}
    \begin{split}
    \dot\rho_{11} =&  - \Gamma_{rel}\lp t \rp sin^2\theta  \rho_{11} + \Gamma_{ex} \lp t \rp sin^2\theta \rho_{22} \\&
                  - \frac{1}{2}sin(2\theta)\left[ \Gamma_{ns}^x\lp t \rp\lp \rho_{12} + \rho_{21} \rp - i\Gamma_{ns}^y\lp t \rp\lp \rho_{12} - \rho_{21} \rp\right]  \\
    \dot\rho_{12} =&  -i\lp E_1 - E_2\rp \rho_{12}  - 2\Gamma_{d}\lp t \rp cos^2\theta\rho_{12} + 2\Gamma_{nr}^+ \lp t \rp sin^2\theta\rho_{21} \\&
                  - \frac{1}{2} sin^2\theta \lp \Gamma_{rel}\lp t \rp \rho_{12} + \Gamma_{ex}\lp t \rp \rho_{12} \rp \\&
                  - \frac{1}{2}sin(2\theta)\left[ \Gamma_{ns}^x\lp t \rp\lp \rho_{11} - \rho_{22} \rp + i\Gamma_{ns}^y\lp t \rp \lp \rho_{11} - \rho_{22} \rp\right]
    \end{split}
\end{array}
\ee
with  $\dot{\rho}_{22} = - \dot{\rho}_{11}$, $\dot{\rho}_{21} = \dot{\rho}_{12}^*$ and $\theta$
as in eq. (\ref{basis}). Details of the derivation as well as explicit expressions for the
coefficients are given in the appendix. (Small) Lamb-shifts have been omitted. This system of
equations contains the standard terms found in its secular approximation, namely, population
relaxation, $\Gamma_{rel}$, and thermal excitation, $\Gamma_{ex}$, which are related by a
detailed balance condition, $\Gamma_{ex}\lp t \rp = e^{-\frac{ \omega_{0} }{kT}}\Gamma_{rel}\lp t \rp$,
and a pure dephasing term with rate ($\Gamma_{d}$). We have retained the time-dependence of
all rates, as we are interested in the early time dynamics. In addition to these contributions,
the full Redfield ME contains additional terms that couple coherences to populations, and coherences
to their complex conjugates. We denote these as non-secular  ($\Gamma_{ns}^{x,y}$) and counter-rotating
\cite{BreuerPetruccione} (or rapidly varying \cite{Blum1981}) ($\Gamma_{nr}^\pm$), respectively.
By counter-rotating we understand those $\Gamma$'s having a time-dependence which oscillates at
twice the frequency $\omega_{0}$ of the excitonic system (these terms would average to zero in
a coarse-grained or rotating-wave approximation) \cite{BreuerPetruccione}. Non-secular terms
have a time-dependence the contains both rotating (slow) \emph{and} counter-rotating (fast) components.

\mbp{For the initial state $\ket{\psi} = \frac{1}{\sqrt{2}}\lp e^{i\xi}\ket{1} + \ket{2}\rp$,
the requirement
$\dot\rho_{11}\lp 0 \rp > 0$ and a Taylor expansion of the decay rates around $t = 0$;
$\Gamma \lp t \rp = \Gamma' \lp 0 \rp t  + O \lp 0 \rp t^2$ leads to the condition
\be
    - \sqrt{2}sin(2\theta) \sin(\xi + \frac{\pi}{4})  > sin^2\theta \, \frac{\Gamma_{rel}\lp 0 \rp}{\Gamma_{ns} \lp 0 \rp}
    \lp 1 - e^{-\frac{\omega_{0}}{kT}} \rp
    \label{inequality}
\ee
for population inversion to occur. In the absence of non-secular terms, $\dot\rho_{11}\lp 0
\rp < 0$ and population inversion cannot occur. The positivity of the right hand side of eq.
\ref{inequality} imposes an initial phase $\xi \in [3\pi/4,7\pi/4]$ for population inversion
to occur, thereby explaining the results of the TEDOPA simulations.} Defining the rate of change
of energy in the \mbp{excitonic subsystem} as $\Delta E(t) = \omega_{0} \dot{\rho}_{11}(t)$,
we observe that the population inversion requires an initial positive $\Delta E(0)$; consequently
there is a net increase of the energy in the system that is being \mbp{provided by} the
non-equilibrium environment.
\mbp{Due to the detailed balance condition on the RHS of Eq. \ref{inequality}, the increase
of the maximum population inversion with temperature, seen in Fig. \ref{Figure2}(b), is neatly
rationalised by the ME analysis.} Finally, we see from Eq. \ref{inequality} that the
counter-rotating terms do not \mbp{play} a role in generating the population inversion,
and we will not discuss these terms any further.

The quantitative agreement \mbp{of the analysis above} with the main trends seen in the
TEDOPA results suggests that a microscopic understanding can be gained by analyzing the
structure of the non-secular terms in Eq. (\ref{reducedequations}). In the excitonic basis,
all dynamics arise from the environment interactions, described through $\mathcal{H_I} =
( g \cos\theta \,\sigma_z - g \sin\theta \,\sigma_x ) \otimes (b_k + b_k^\dagger),$ where
$b_k$ denote environmental operators.

\mbp{The excitonic coupling to the environment therefore includes a longitudinal (relaxation)
component in addition to the transversal (dephasing) noise that lead to interference effects
and the dynamical coupling of populations and coherences.} Non-secular terms in the time evolution
can no longer be neglected, as the resulting quantum interference will be crucial to understand
the dynamics in the ultrafast time scale.

Heuristically, the physical origin of the phase-dependence of excitonic transport can be
understood by considering the transition amplitudes between the initial state $\ket{\psi}
=\frac{1}{\sqrt{2}}\lp e^{i\xi}\ket{1} +  \ket{2}\rp$ and the eigenstates $\ket{1},\ket{2}$.
A direct calculation yields $\bra{1}|H_{I}\ket{\psi} = \frac{1}{\sqrt{2}}\left[\sin(\theta)
-e^{i\xi}\cos(\theta)\right]X_{f_{b1}i}$ and $\bra{2}|H_{I}\ket{\psi} = \frac{1}{\sqrt{2}}
\left[\sin(\theta)+e^{i\xi}\cos(\theta)\right]X_{f_{b2}i}$, where the environment matrix
element is between initial and final bath configurations. These amplitudes can be understood
simply as arising from two interfering pathways; \mbp{the amplitude for a flip
$|1\rangle\leftrightarrow |2\rangle$ being proportional to $\sin(\theta)$ and the amplitude
for the populations to remain unchanged, $|i\rangle\rightarrow|i\rangle$, which is
proportional to $\cos(\theta)e^{i\xi}$.}  The key observation is that $\xi$ controls whether
the interference is constructive or destructive for a given transition (conservation of
probability ensures that the other transition is suppressed or enhanced, accordingly). \mbp{Moreover,
by virtue of the $sin(2\theta)$ proportionality of the non-secular terms the mixing angle
$\theta = \pi/4$ maximises the  population inversion, corresponding to a maximum interference
between the longitudinal and transversal components of the environment while completely delocalised
($\theta = \pi/2$) or localised ($\theta = 0$) states lead to no inversion in the population
evolution.} The interference \mbp{between population preserving and inverting} pathways is
described in the ME approach by the non-secular terms. Those terms are characterised for not
conserving the energy in the system. The physical framework is thus that of an electronic
transition in the system being compensated \mbp{for} by the creation/anihilation of a virtual phonon
in the environment. This is a process that drives the environment out of equilibrium, and
consequently must vanish in the longer time-scales, thereby setting the transient nature of
the population inversion and ensuring always the relaxation towards a thermal state.

Our ME analysis demonstrates that the phase-dependent transport uncovered by \mbp{the TEDOPA
numerics is not a peculiarity of a specific} set of parameters but a rather general feature
of composite open quantum systems, with the magnitude of the population inversion \mbp{being}
sensitive to the environment spectral function and the mixing angle \mbp{as determined} by the
coherent coupling between sites. Indeed, we can reproduce the qualitative features of the effect
observed by TEDOPA employing the Redfield ME. These results, presented in the S.I. also show that
any violation of positivity (a known problem of equations beyond the Born-Markov approximation
\cite{Davies1974,BreuerPetruccione}) of the reduced density matrix is several orders of magnitude
smaller in magnitude than the \mbp{observed population inversion and this together with the
fact that TEDOPA generates, by construction, a manifestly positive and numerically exact many-body
density matrix, indicate that this effect is real and observable in principle.}

\begin{figure}[!h]
\includegraphics[width=7cm]{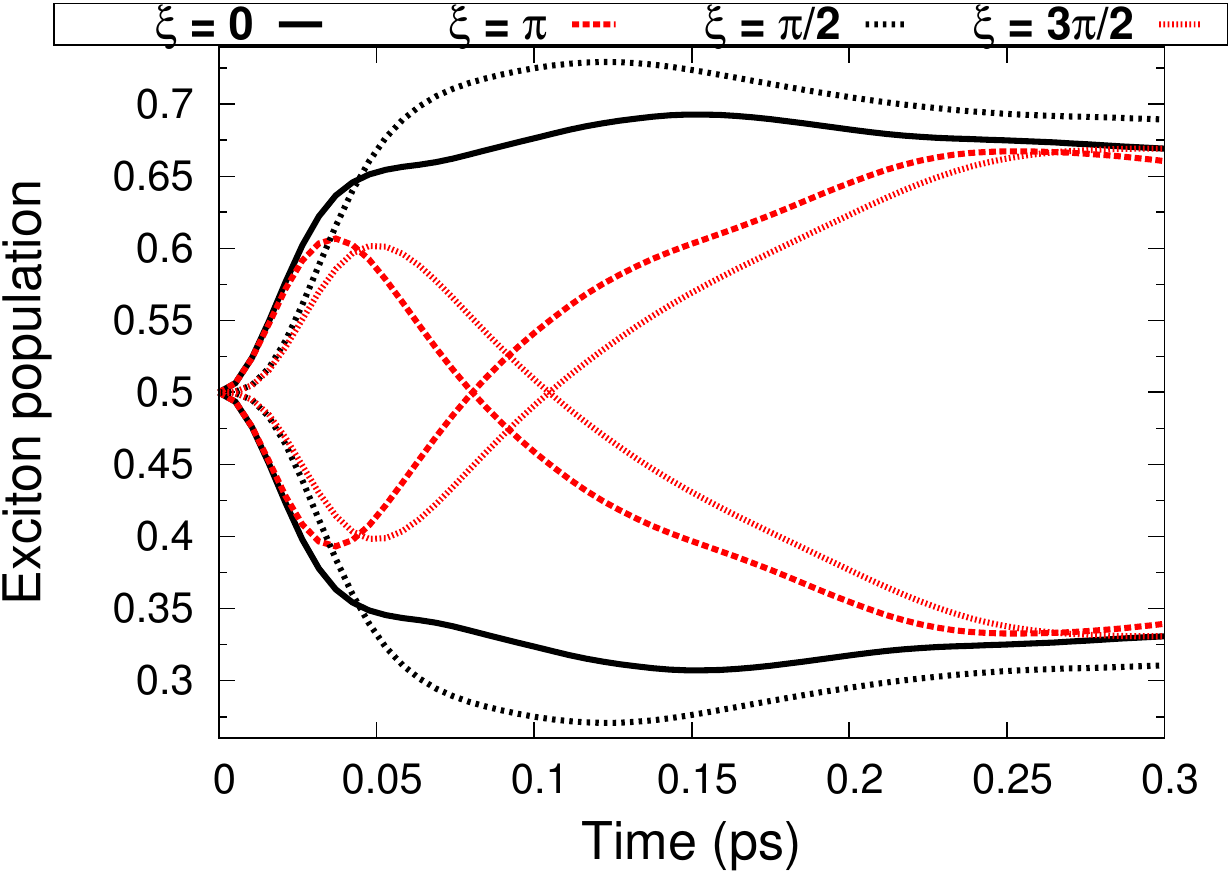}
\caption{TEDOPA numerical results for the time evolution of exciton eigenstate populations
on a dimer with $\omega_a - \omega_b = 135 \, cm^{-1}$ and $J = 52 \, cm^{-1}$ in the
superohmic background charaterised by the AR spectral density with two discrete vibrational
modes of frequencies 180 $cm^{-1}$ and 37 $cm^{-1}$ at T = 277K for representative
$\xi = 0, \pi/2, \pi, 3\pi/2$  phases.}
\label{Figure5}
\end{figure}

Returning to the example of PPCs, most realistic spectral functions contain, in addition to the
smooth background, a number of sharp features, usually associated with underdamped intramolecular
vibrational modes \cite{prior2010efficient,ChinHuelgaPlenio2012,chin2013role,kolli2012fundamental,Womick2011}.
We tackle this more complicated problem with TEDOPA, and examine the influence of two such modes
(one of them resonant with the excitonic gap) in the phase-dependent population inversion. Fig.
\ref{Figure5} shows that the maximum inversion is significantly enhanced (up to 20 $\%$ for T = 277 K
and $\xi = \pi$ as compared to Fig. \ref{Figure2}[a]) by the presence of discrete modes. The situation
is easily understood via an \emph{exact} quantum mechanical analysis of the system \emph{and} the
environment in the limit t $\rightarrow$ 0. With factorised initial conditions, the second derivative
of the high-energy exciton population at time zero is given by:

\be
\label{QManalysis}
\begin{split}
\ddot \rho_{11}\lp 0 \rp &=-\sin(2\theta)\Re\left\{\rho_{e_{1}e_{2}} (0)\right\}\int_{0}^{\infty}S_{0}(\omega) (2n(\omega)+1)d\omega  \\&+ \sum_{i}g_{i}^2(2n(\omega_{i})+1),
\end{split}
\ee
where $n(\omega)$ is the Planck distribution, $S_0 \lp k \rp$ is the continuum part of
the spectral function $S(\omega)$ and $i$ represent an arbitrary finite number of intramolecular
modes that couple to the system with amplitude $g_{i}$. As it can also be shown that
$\dot\rho_{11}(0)=0$ (see SI), Eq. \ref{QManalysis} shows that discrete modes will always
increase the initial rate of energy flow, explaining the enhancement of the population
inversion seen in Fig. \ref{Figure5}. Interestingly, the analysis shows that the initial
enhancement due to discrete modes does not depend on the frequency of the modes, i.e.
exact resonance is not required (frequency only appears through the temperature dependence).

$\textit{Conclusions}$ - By combining powerful numerical techniques with theoretical model
analysis, we have demonstrated that electronic coherence in a dimeric system can not only
quantitatively influence the flow of energy transfer over the correlation time of the
environment, it may even revert the \emph{direction} of the flow and permit transient energy
harvesting from the surroundings. By virtue of the Redfield equation analysis, we were able
to predict the electronic and environ- mental properties that maximise this effect, and also
showed that - in principle - such effects could be found in a wide range of natural dimeric
systems. As just one possible example, the experimental preparation of different excited state
coherences might be achievable in molecular systems via polarisation control or quantum control
techniques in 2D Fourier Transform spectroscopies. These methods have already been shown to be
capable of following dynamics within the typical bath correlation times we have consider here
$(\approx 50-100 fs)$ \cite{engel2007evidence,Fleming2008,collini2010coherently,romero2014quantum,Jaemin2015}.
Finally,
we remark that the fundamental processes we describe
strongly motivate the design of thermal energy harvesting in
quantum devices. As it is known from classical systems, these
transient effects need to rectified in order to be used, which
presents a rather interesting theoretical problem in the context
of multi-component, non-equilibrium quantum dynamics.

\begin{acknowledgments}
We are most grateful to Andrea Smirne for fruitful discussion and critical reading of the manuscript.
This work was supported by an Alexander von Humboldt Professorship, the EU STREPs PAPETS and
QUCHIP, the EU Integrating Project SIQS and the ERC Synergy grant BioQ. AWC acknowledges
support from the Winton Programme for the Physics of Sustainability. JP acknowledges funding by the Spanish
Ministerio de Econom\'ia y Competitividad under Project No.
FIS202-3062.
\end{acknowledgments}

\bibliography{./PDTbib.bib}

\clearpage

\appendix
\label{SI}

\subsection{Dimer parameters and Spectral function for TEDOPA simulations}

TEDOPA simulations were carried out for a dimer system with local disorder
$\omega_a - \omega_b = 135 \, cm^{-1}$ and coherent coupling $J = 52 \, cm^{-1}$,
yielding a mixing angle of $\theta = \pi/4$. These parameter choice is motivated
by an actual natural system as they correspond to the two lowest sites energies
of the FMO photosynthetic complex and their electrostatic coupling \cite{adolphs2006proteins}.
This model system has been previously considered in the discussion of long lived
excitonic coherence \cite{chin2013role}.

The spectral function used to characterise the system-environment interaction in
TEDOPA simulations is given by
\be
\begin{split}
S\lp \omega \rp &= \frac{\lambda \left[ 1000 \omega^5 e^{-\sqrt{\frac{\omega}{\omega_1}}} +
4.3 \omega^5 e^{-\sqrt{\frac{\omega}{\omega_2}}}\right] }{9!\lp1000 \omega_1^5 + 4.3 \omega_2^5\rp} \\& + \sum_{i = 1}^2 g_i \omega_i^2 \delta(\omega - \omega_i).
\end{split}
\ee

The parameters for the continuum part are $\lambda = 35 cm^{-1}$, $\omega_1 = 0.57 cm^{-1}$
and $\omega_2 = 1.9 cm^{-1}$ as taken from the experimental fitting provided in \cite{adolphs2006proteins}.
This smooth background differs from the standard Drude-Lorentz function in being super-Ohmic,
with the high frequency part of the spectrum responsible for fast thermalization, as required
for efficient energy transfer.

The considered discrete modes have frequencies $\omega_1 = 180 cm^{-1}$ and $\omega_2 = 37 cm^{-1}$
and couplings $g_1 = 0.12$, $g_2 = 0.22$, as reported in \cite{vibrationalmodesFMO,adolphs2006proteins}.


\subsection{Master equation derivation}

Our theoretical model is provided by an excitonically coupled dimer (sites a and b) with Hamiltonian
\be
{\mathcal{H}}_S = \frac{\omega_a}{2}{\sigma}_z^a + \frac{\omega_b}{2}e{\sigma}_z^b + J\lp {\sigma}_+^a{\sigma}_-^b + {\sigma}_-^a{\sigma}_+^b \rp.
\ee

Each site is coupled to a vibrational environment modelled as an infinite bath of harmonic oscillators, whose Hamiltonian is ${\mathcal{H}}_B = \sum_i \int_0^{\infinity}
dkg_i\lp k \rp b_i^\dagger \lp k \rp b_i \lp k \rp$. The interaction Hamiltonian describing the electron-phonon coupling is considered to be of the form
\be
{\mathcal{H}}_I = \sum_i \lp \sigma_z^i + I \rp \int_0^{\infinity} dk h_i(k) \lp b_i(k) + b_i^{\dagger}(k) \rp.
\ee

We are interested in the time evolution of the diagonal states of ${\mathcal{H}}_S$, known
as exciton states. Moreover, we will ignore excitonic states that correspond to either no
excitation on each site ($\ket{g_ag_b}$), or to both sites excited ($\ket{e_ae_b}$), where
\mbp{the former are irrelevant to our discussion and the later are typically strongly suppressed
in the systems of interest \cite{Renger2001}.} Hence, working on the one excitation sector,
spanned by the (localized) states  $\ket{e_ag_b}$ and $\ket{g_ae_b}$, the excitonic states
can be expressed a coherent superposition of the form
\be
 \left(
\begin{array}{c}
\ket{1} \\
\ket{2}
\end{array}
\right) =
\left(
\begin{array}{cc}
cos\frac{\theta}{2} & \sin\frac{\theta}{2} \\
-sin\frac{\theta}{2} & \cos\frac{\theta}{2}
\end{array}
\right)
\left(
\begin{array}{c}
\ket{e_ag_b} \\
\ket{g_ae_b}
\end{array}
\right),
\label{basis2}
\ee

with energies $E_{1,2} = \pm \frac{1}{2}\sqrt{4J^2 + \Delta^2} = \pm \frac{\omega_0}{2}$,
where $\Delta = \omega_a - \omega_b \, \lp\omega_a > \omega_b\rp$ and $\tan \,\theta = 2J/\Delta$
is the so called mixing angle, a measure of the degree of delocalization resulting from the
coherent coupling $J$. The excitonic splitting is then given by $\omega_0$. At this point it
is convenient to redefine the environmental bosonic operators as $b_{p,m} =
\frac{1}{\sqrt{2}}\lp b_a \pm b_b \rp$, which when we impose the restriction of one excitation
sector, will remove the contribution from the $b_p$ operators, thus leaving just one set of
harmonic oscillators. The complete Hamiltonian for the effective global system (excitonic +
vibrational degrees of freedom) is then $\mathcal{H}_S + \mathcal{H}_B + \mathcal{H}_I$, where
\be
    \begin{split}
    &\mathcal{H}_S = \frac{1}{2}\omega_{0} \sigma_z  \\&
    \mathcal{H}_B = \int_0^{\infinity} dk  h(k)b^{\dagger}(k)b(k)  \\&
    \mathcal{H}_I = \lp cos\theta\sigma_z + sin\theta\sigma_x \rp \int_0^{\infinity}dk g(k) \lp b(k) + b^{\dagger}(k) \rp.
    \end{split}
\ee
For simplicity we have assumed that both sites are subject to the same type of environment.

In order to obtain the Master Equation for the evolution of the excitonic system, we move to
an interaction picture with respect to the Hamiltonian $\mathcal{H} = \mathcal{H}_S + \mathcal{H}_B $,
in which,
\be
    A^I \lp t \rp = \exp \lp i \mathcal{H}t \rp A \exp \lp - i \mathcal{H}t \rp.
\ee
From now on we omit the superscript ``I'' on the interaction picture operators.
Assuming a factorized initial state of the form $\rho \lp t \rp = \rho_S \lp t \rp\otimes \rho_B$
and working in the Born-Markov approximation, the von-Neumann equation for the evolution
of the density matrix reads
\be
\dot\rho \lp t \rp = - \int_0^t dt_1 Tr_B \left\{ \left[ \mathcal{H}_I \lp t \rp, \left[ \mathcal{H}_I \lp t_1 \rp, \rho_S \lp t \rp \otimes \rho_B \right]\right] \right\}
\ee
which is a time local Master Equation, also known as the Redfield equation
\cite{BreuerPetruccione}. The coupling Hamiltonian $H_I (t)$ in the interaction
representation is given by
\be
\begin{split}
\mathcal{H}_I =& \left[ cos\theta\sigma_z - sin\theta \lp cos\omega_{0} t \sigma_x - sin\omega_{0} t \sigma_y \rp \right] \\&
\otimes \int_0^{\infinity} dk \lp b(k) e^{-i\omega t} + b(k)^\dagger e^{i\omega t} \rp,
\end{split}
\ee
where we have redefined the bath operators as the product $g\, b(k)$ and assumed the
coupling $g$ to be $k$-independent. The presence of crossed terms involving distinct
Pauli operators will lead to non-secular contributions in the resulting excitonic master
equation. To perform the derivation, the following identities prove useful
\be
\begin{split}
& \sigma_j = \left(
\begin{array}{cc}
\delta_{jz} & \delta_{jx} - i\delta_{jy}\\
\delta_{jx} + i\delta_{jy} & \delta_{jz}
\end{array}
\right) \\&
\left[ \sigma_i,\sigma_j \right] = 2i\epsilon_{i,j,k}\sigma_k \\&
\left\{ \sigma_i,\sigma_j \right\} = 2\delta_{i,j} I \\&
\sigma_x = \sigma_- + \sigma_+ \, \,, \, \, \sigma_y = i\lp \sigma_- - \sigma_+ \rp.
\end{split}
\ee

Disregarding Lamb shifts, which are unimportant in our context, the time-local Master
Equation for the excitonic dynamics can be written as follows
\begin{widetext}
\be
\label{completeME}
\begin{split}
\frac{d}{dt}\rho \lp t \rp =& - \Gamma_{d}\lp t \rp cos^2\theta \lp \rho \lp t \rp  -   \sigma_z \rho \lp t \rp \sigma_z \rp
                    + sin^2\theta \lp \Gamma_{nr}^+\lp t \rp\sigma_+ \rho \lp t \rp \sigma_+ + \Gamma_{nr}^-\lp t \rp\sigma_- \rho \lp t \rp \sigma_- \rp \\&
                    + \Gamma_{rel}\lp t \rp sin^2\theta  \lp\sigma_-\rho \lp t \rp \sigma_+ - \frac{1}{2}\left\{ \sigma_+\sigma_-,\rho \lp t \rp \right\}\rp
                    + \Gamma_{ex}\lp t \rp sin^2\theta  \lp\sigma_+\rho \lp t \rp \sigma_- - \frac{1}{2}\left\{ \sigma_-\sigma_+,\rho \lp t \rp \right\}\rp \\&
                    - \frac{1}{2}sin(2\theta) \lp\Gamma_{ns}^x \lp t \rp  \lp \sigma_z \rho \lp t \rp  \sigma_x + \sigma_x   \rho \lp t \rp \sigma_z \rp
                    - \Gamma_{ns}^y \lp t \rp  ( \sigma_z \rho \lp t \rp  \sigma_y + \sigma_y   \rho \lp t \rp \sigma_z ) \rp,
\end{split}
\ee
\end{widetext}

where the expressions for the different rates are given by

\be
\begin{split}
& \Gamma_{d} \lp t \rp = 2\int_0^t dt_1 Q \lp t - t_1 \rp, \\&
\Gamma_{nr}^+ \lp t \rp = 2\int_0^t dt_1 e^{i\lp \omega_{0}\lp t + t_1 \rp \rp}    Q \lp t - t_1 \rp, \\&
\Gamma_{nr}^- \lp t \rp = 2\int_0^t dt_1 e^{-i\lp \omega_{0}\lp t + t_1 \rp \rp}    Q \lp t - t_1 \rp, \\&
\Gamma_{rel} \lp t \rp = 4\int_0^t dt_1 cos \lp \omega_{0}\lp t - t_1 \rp \rp Q \lp t - t_1 \rp,     \\&
\Gamma_{ex} \lp t \rp = 4\int_0^t dt_1 cos \lp \omega_{0}\lp t - t_1 \rp \rp e^{-\omega_{0}\beta } Q \lp t - t_1 \rp,   \\&
\Gamma_{ns}^x \lp t \rp = 2\int_0^t dt_1  cos\frac{\omega_{0}\lp t - t_1 \rp}{2}cos\frac{\omega_{0}\lp t + t_1 \rp}{2} Q \lp t - t_1 \rp, \\&
\Gamma_{ns}^y \lp t \rp = 2\int_0^t dt_1  sin\frac{\omega_{0}\lp t - t_1 \rp}{2}sin\frac{\omega_{0}\lp t + t_1 \rp}{2} Q \lp t - t_1 \rp.
\end{split}
\ee

The function $Q \lp t - t_1 \rp$ gathers the effect of the environment as
\be
Q \lp t - t_1 \rp = \int_0^{\infty} d\omega \, S\lp \omega \rp coth \lp \frac{\beta \omega}{2}\rp cos\lp \omega \lp t - t_1 \rp \rp,
\ee
with the spectral density $S(\omega)$ describing the weighted distribution in
frequency space of the environment of harmonic oscillators.



The dynamics resulting from the considered ME approach is illustrated in Fig. 4, where we show the time evolution of the excitonic populations and the rate of change of energy, as well as a positivity test of the resulting (non-secular) density matrix. In these results we
introduced heuristically a decay for the non-rotational and non-secular rates via the time dependent exponential $exp\left\{-\omega_{0} t \right\}$, while for simplicity leaving the rest
of the rates equal and constant.

\begin{figure}[ht!]
\includegraphics[width=\columnwidth]{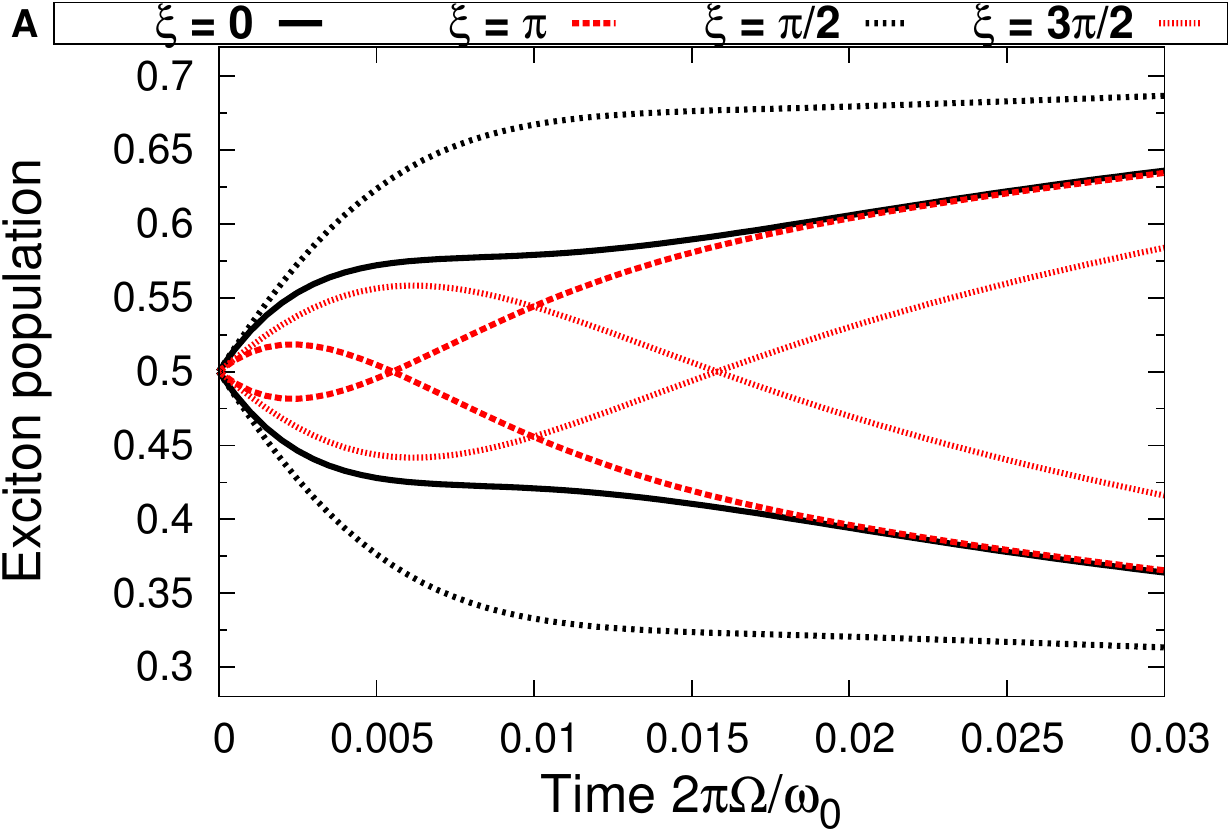}
\includegraphics[width=\columnwidth]{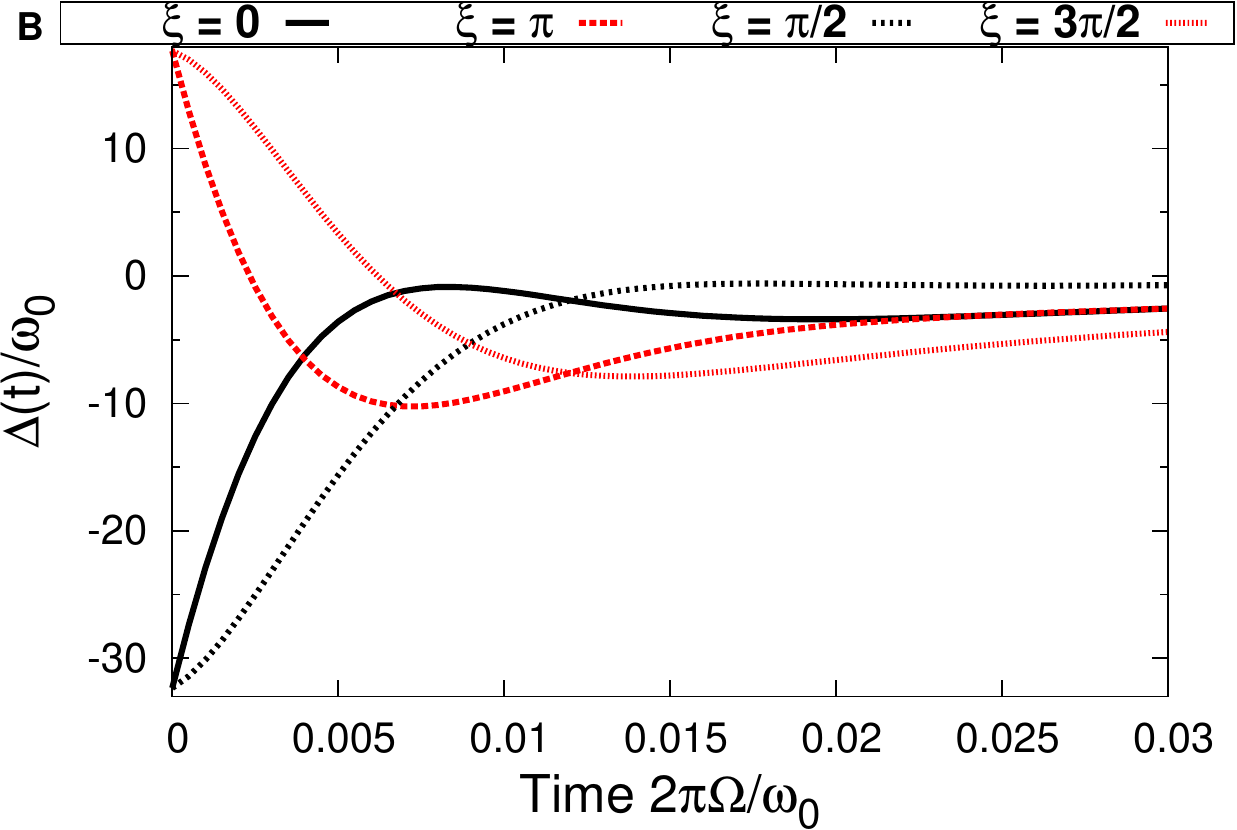}
\includegraphics[width=\columnwidth]{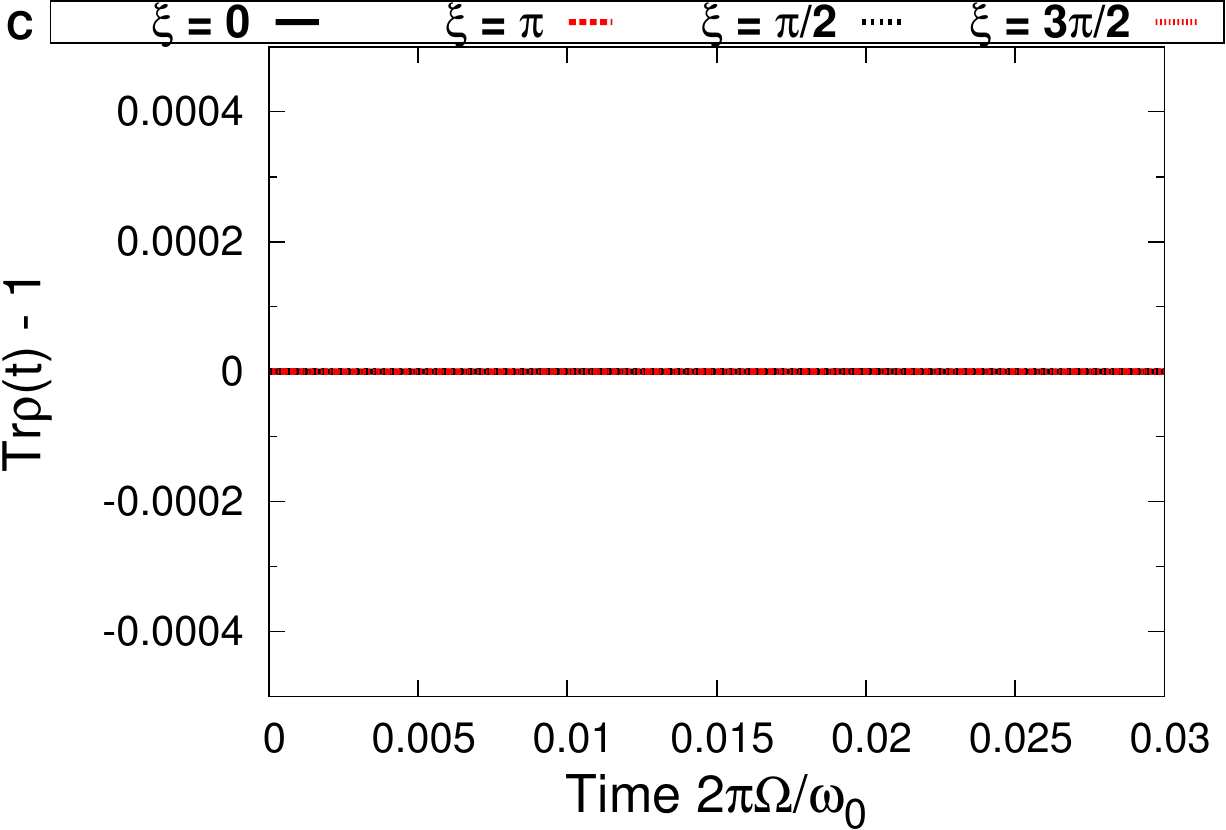}
\caption{Master Equation simulation (a) Excitonic eigenstate population obtained by numerically solving the ME Eq. \ref{completeME} for representative $\xi = 0, \pi/2, \pi, 3\pi/2$ at T = 277K. Note that the crossing
is qualitatively reproduced even in this simple situation. (b) Rate of change of energy $\Delta E(t) =  \dot{\rho}_{11}(t) $ for representative $\xi = 0, \pi/2, \pi, 3\pi/2$ at T = 277K, positive
the differences when the crossing is present, and always asymptotically going to zero as required by returning to equilibrium. (c) Positivity test for representative $\xi = 0, \pi/2, \pi, 3\pi/2$ at T = 277K. Although
it is not ensured, note that the scale of violations would be orders of magnitude smaller than the excitonic difference in the crossing on (a).}
\label{modelpopulations}
\end{figure}



\subsection{Exact derivation with discrete modes}

Let us analyse the effect of the presence of intramolecular vibrational modes
by performing an exact quantum mechanical calculation, which \mbp{is} possible
at $t = 0$ when assuming a factorized initial condition. Using the excitonic
basis for the dimeric system, we can write
\be
    \mathcal{H} = \sum_{i = 1,2} E_{n} \ket{n}\bra{n} + \sum_{i,k} \omega_k c_{i,k}^\dagger c_{i,k}
    + \sum_{n,m} Q_{nm} \ket{n}\bra{m}.
\ee
Here we have assumed a finite number $k$ of independent bath modes coupled to
each site of the dimer. In the delocalised exciton basis the coupling system-environment
gets redefined through the new couplings $Q_{nm} = \sum_{i,k} g_{i,k} C_n^i C_m^i ( c_{i,k} + c_{i,k}^\dagger )$, where the coefficients C can be taken as real, so $Q_{nm} = Q_{mn}$. These coefficients
can be re-expressed in terms of the mixing angle of the dimer system to obtain explicit
expressions for the Q couplings
\be
\begin{array}{c}
\begin{split}
Q_{11} &= \sum_k g_k cos^2 \lp \frac{\theta}{2} \rp \lp c_{1,k} + c_{1,k}^\dagger \rp \\& + \sum_k g_k sin^2 \lp \frac{\theta}{2} \rp \lp c_{2,k} + c_{2,k}^\dagger \rp
\end{split} \\
\begin{split}
Q_{22} &= \sum_k g_k sin^2 \lp \frac{\theta}{2} \rp \lp c_{1,k} + c_{1,k}^\dagger \rp \\& + \sum_k g_k cos^2 \lp \frac{\theta}{2} \rp \lp c_{2,k} + c_{2,k}^\dagger \rp
\end{split} \\
\begin{split}
Q_{12} &= - \frac{1}{2}\sum_k g_k sin \lp \theta \rp \lp c_{1,k} + c_{1,k}^\dagger \rp \\& + \frac{1}{2}\sum_k g_k sin \lp \theta \rp \lp c_{2,k} + c_{2,k}^\dagger \rp.
\end{split}
\end{array}
\ee
Introducing the mode displacements $X_k = c_{k} + c_{k}^\dagger$ allows us to write
\be
\begin{split}
    &Q_{11} - Q_{22} = cos\theta \sum_k g_k \lp X_{1,k} - X_{2,k} \rp \\&
    Q_{12} = -\frac{sin\theta}{2} \sum_k g_k \lp X_{1,k} - X_{2,k} \rp.
\end{split}
\ee

Using the Heisenberg equations of motion $\dot{O}(t) = -i \left[\mathcal{H},O(t)\right]$ to
obtain the equations of motion for the excitonic population and coherence operators we obtain
\be
\begin{split}
&\dot{\rho}_{11} \lp t \rp = -i Q_{12}\lp t \rp \lp \rho_{12}\lp t \rp - \rho_{21}\lp t \rp \rp  \\&
 \dot{\rho}_{12} \lp t \rp = -i\lp E_{12} + Q_{11} \lp t \rp - Q_{22} \lp t \rp \rp \rho_{12} \lp t \rp  \\& \,\,\,\,\,\,\,\,\,\,\,\,\,\, -i Q_{12} \lp t \rp \lp \rho_{22}\lp t \rp - \rho_{11}\lp t \rp \rp,
\end{split}
\ee
Following the analysis performed on the main text, we want to know the influence that a phase of a superposition of excitons
has on the direction of the energy flow. The rate of change of energy is
$\Delta(t) = \omega_0 \dot\rho_{11}(t)$, so the sign of $\rho_{11}(t)$ at
$t = 0$ tells in which direction the energy flows after a sharp excitation.
Thus, taking the second derivative of the high energy exciton population
evolution equation and inserting the expression above for the rate of change
of coherences we obtain
\be
\ddot\rho_{11}\lp t \rp =  Q_{12}\lp 0 \rp \lp E_{12} + Q_{11}\lp 0 \rp - Q_{22}\lp 0 \rp \rp \Re\left\{\rho_{12}\lp 0 \rp \right\}.
\ee

\begin{figure}[h!]
\includegraphics[width=\columnwidth]{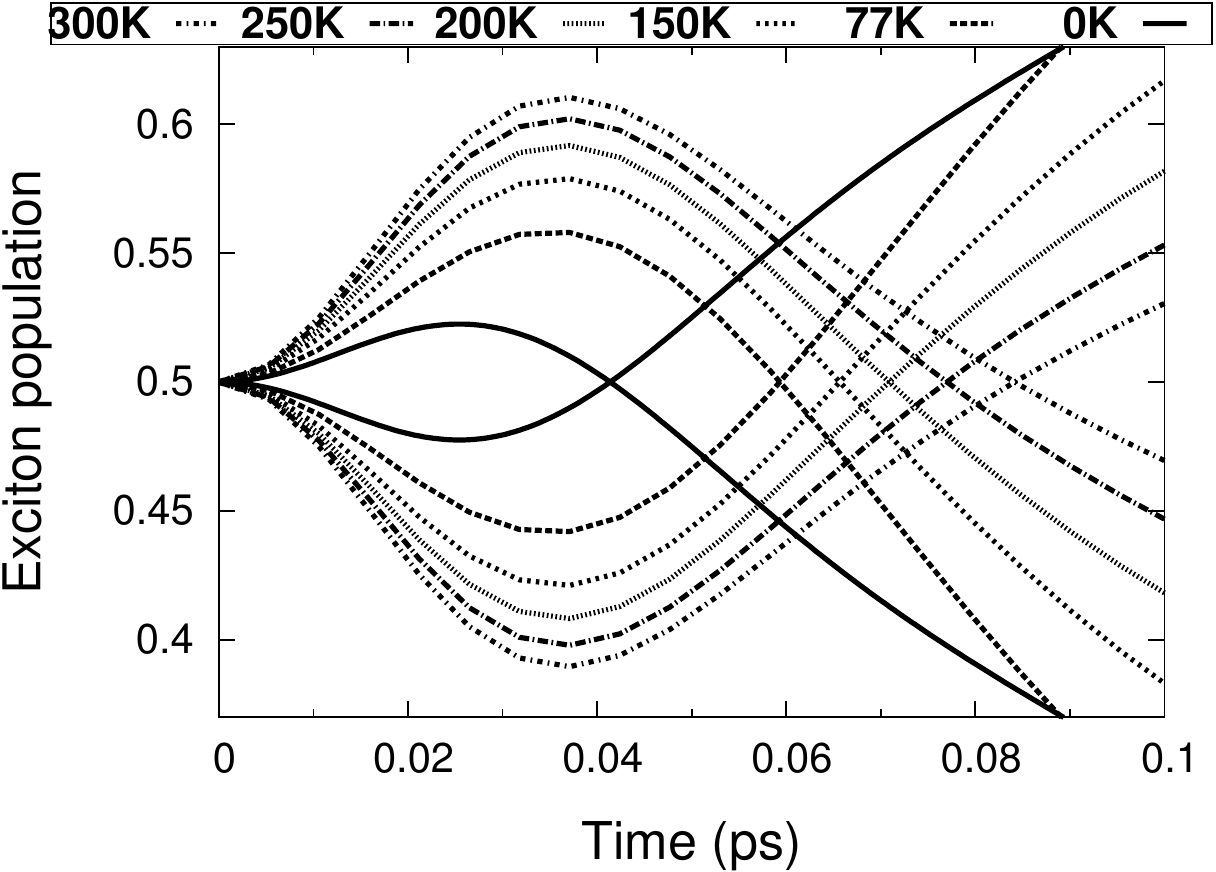}
\caption{TEDOPA numerical results for the time evolution of exciton eigenstate populations on a dimer with $\omega_a -\omega_b = 135 \, cm^{-1}$ and
 $J = 52 \, cm^{-1}$ in the presence of two vibrational modes of frequencies 180 $cm^{-1}$ and 37 $cm^{-1}$ for temperatures $0-300$ K following the initial state with $\xi = \pi$ in the AR background.}
\label{Figure6}
\end{figure}

Assuming separability into electronic and thermal environment density matrices,
the expectation of the second derivative of the population with respect to the
latter is
\be
\ddot\rho_{11}\lp t \rp = 2 \langle Q_{12}\lp 0 \rp (Q_{11}\lp 0 \rp - Q_{22}\lp 0 \rp \rangle_\beta \Re\left\{\rho_{12}\lp 0 \rp \right\}
\ee
where the thermal expectation value of the environment operators is
\be
\langle Q_{12}\lp 0 \rp (Q_{11}\lp 0 \rp - Q_{22}\lp 0 \rp \rangle_\beta = -\frac{1}{2}sin(2\theta)\sum_k g_k^2(2n(\omega_k) + 1),
\ee
with $n(\omega)$ being the thermal \mbp{Planck} distribution. $\langle Q_{ij} \lp t \rp \rangle$
\mbp{vanishes} for a thermal state (\mbp{explaining} why the first derivative of the populations
vanishes as $t \rightarrow 0$). We can now combine those two last equations to obtain
\be
    \label{expop}
    \frac{d^2\rho_{11}}{dt^2}|_{t=0} = -sin \lp 2\theta \rp \Re\left\{\rho_{12}\lp 0 \rp \right\} \sum_k g_k^2(2n(\omega_k) + 1).
\ee
Introducing now the definition for the spectral function of the environment as $S(\omega)=  \sum_{i}g_i^2 \delta(\omega-\omega_{i})$ allows us to write:

\be
\sum_{i}g_i^2 (2n(\omega_i) + 1) = \int_0^{\infty} S \lp \omega \rp \lp 2n\lp\omega\rp + 1 \rp d\omega.
\ee

And if we now choose the spectral density to contain discrete vibrational modes as $S(\omega)= S_0(\omega) + \sum_{i}g_k^2 \delta(\omega-\omega_{i})$, with
$S_0\lp \omega \rp$ a smooth function representing the background fluctuations and $i$ discrete modes of frequency $\omega_i$ and coupling strength $g_i$, we obtain
\be
\label{SpectralFunction}
\begin{split}
&\sum_{k}g_k^2(2n(\omega_{k})+1) = \int_{0}^{\infty}S(\omega)(2n(\omega)+1)d\omega = \\&
\int_{0}^{\infty}S_{0}(\omega) (2n(\omega)+1)d\omega  + \sum_{i}g_{i}^2(2n(\omega_{i})+1).
\end{split}
\ee

Putting together this result and the equation for the second derivative of the exciton population Eq- \ref{expop}, we observe that the presence of discrete modes will
always enhance the initial rate of energy flow, whichever its direction.

We recover here the result that the phase sensitivity is introduced in the system via a correlation of transversal and longitudinal components of the environment
 ($\langle Q_{12}\lp 0 \rp (Q_{11}\lp 0 \rp - Q_{22}\lp 0 \rp \rangle$) in Eq \ref{expop}, in complete agreement with the ME approach presented above,
where it is also obtained that the optimal mixing angle $\theta = \pi/4$ maximises the effect (overlap) of these cross correlations. In addition, the quantum mechanical analysis, although valid only at t = 0 (afterwards
no longer can there be factorisation), shows that the populations grow quadratically at early times, as obtained with TEDOPA.


\end{document}